\pdfoutput=1
\documentclass[bib]{statapress}

\usepackage[crop,newcenter,frame]{pagedims}

\usepackage{sj}

\usepackage{epsfig}

\usepackage{stata}

\usepackage{shadow}

\usepackage{amsmath}

\sjsetissue{$vv$}{$ii$}{$mm$}{$yyyy$}

\begin{document}


\inserttype[st0001]{article}
\author{Grayling, Wason, and Mander}{%
  Michael J. Grayling\\Hub for Trials Methodology Research\\MRC Biostatistics Unit\\Cambridge, UK\\mjg211@cam.ac.uk
  \and
  James M. S. Wason\\Hub for Trials Methodology Research\\MRC Biostatistics Unit\\Cambridge, UK\\
  Institute of Health \& Society\\Newcastle University\\Newcastle, UK\\james.wason@mrc-bsu.cam.ac.uk
  \and
  Adrian P. Mander\\Hub for Trials Methodology Research\\MRC Biostatistics Unit\\Cambridge, UK\\adrian.mander@mrc-bsu.cam.ac.uk
}
\title[Group Sequential Trial Design]{Group Sequential Clinical Trial Designs for Normally Distributed Outcome Variables}
\maketitle

\begin{abstract}
In a group sequential clinical trial, accumulated data are analysed at numerous time-points in order to allow early decisions about a hypothesis of interest. These designs have historically been recommended for their ethical, administrative and economic benefits. In this work, we discuss a collection of new Stata commands for computing the stopping boundaries and required group size of various classical group sequential designs, assuming a normally distributed outcome variable. Following this, we demonstrate how the performance of several designs can be compared graphically.
\end{abstract}

\section{Introduction}

Parallel group randomised controlled trials are typically conducted by recruiting a fixed number of individuals and allocating each to receive one of two treatments, ultimately testing a pre-specified hypothesis. Since Wald published his work on the sequential probability ratio test \citep{wald1947}, there has been substantial interest in trial designs that allow hypotheses to be tested multiple times during the trial. With this approach, the trial may be stopped early if the data so suggests. This leads to patient exposure to inferior treatments being limited, and, by helping to lower the expected required sample size, the cost of a trial will often be reduced.

\citet{Armitage1975} was responsible for much of the early use of such methods in medicine. However, his and the other initial approaches were fully sequential, with data analysed after every patient. Whilst this may seem desirable, it is impractical and thus this methodology did not gain general acceptance. The pivotal moment in this field came with the work of \citet{pocock1977}, who provided a clear way of determining group sequential designs with desired type-I and type-II error rates. In a group sequential design, a hypothesis is analysed multiple times during an on going trial, but as the name suggests, only after groups of certain sizes have been assessed. This allows the majority of the benefits of a fully sequential approach to be retained, whilst also making the design feasible in practice.

Since this paper, group sequential designs have been researched extensively, and utilised regularly in clinical trials. Today, methodology is well established for designing group sequential trials with normal, binary, and survival endpoints. Approaches are available to design trials with unknown variance, with multiple arms, or to optimise a designs features. For a detailed discussion of available methods see \citet{whitehead1997} or \citet{jennison2000}. 

In this paper, we focus on the design of two-treatment group sequential trials with a normally distributed outcome variable, but note that asymptotically other endpoint types can be treated with the same normal test statistics. We proceed by summarising the statistical theory behind group sequential methodology. Following this we detail our new commands, and provide several examples of their use.

\section{Statistical Theory}

We consider a randomised two-arm group sequential trial design with up to $L$ planned analyses. We index one arm by 0, and the other by 1. Often, it will be the case that arm 0 is a control and arm 1 an experimental treatment, but this may not always be true. We assume that the $l$th analyses takes place after $n_{0l}=ln$ and $n_{1l}=rln$ patients have been randomised to arms 0 and 1, respectively. Possible extensions to this framework are discussion in Section 8. Thus $r$ is the ratio of patients allocated to arm 1 relative to arm 0, and we refer to $n$ as the group size. The outcome from patient $i$ in arm $d$ in stage $l$ , $Y_{dli}$, is assumed to be distributed as follows
\[Y_{dli} \sim N(\mu_d, \sigma_d^2).\]
Thus, we are assuming that the variance in response of both treatments is known.

Our ultimate goal is to make inference about the difference in the average treatment effect of arms 0 and 1. To this end, we define $\tau = \mu_1 - \mu_0$, and at each interim analysis $l$ compute the following test-statistic
\[ Z_l = \left(\frac{1}{n_{1l}}\sum_{j=1}^{l}\sum_{i=1}^{rn}Y_{1ji} - \frac{1}{n_{0l}}\sum_{j=1}^{l}\sum_{i=1}^{n}Y_{0ji}\right)I_{l}^{1/2},\]
with
\begin{equation} \label{eq:I}
I_l = \left(\frac{\sigma_0^2}{n_{0l}} + \frac{\sigma_1^2}{n_{1l}} \right)^{-1} = \left(\frac{\sigma_0^2}{ln} + \frac{\sigma_1^2}{rln} \right)^{-1},
\end{equation}
the information for this analysis. It can be shown that $\{Z_1,\dots,Z_L\}$ have for the parameter of interest $\tau$, with information levels $\{I_1,\dots,I_L\}$, what has been referred to as the canonical joint distribution \citep{jennison2000}. That is
\begin{itemize}
	\item $(Z_1,\dots,Z_L)$ is multivariate normal;
	\item $E(Z_l)=\tau I_l^{1/2}$, $l=1,\dots,L$;
	\item $cov(Z_{l_1},Z_{l_2}) = (I_{l_1}/I_{l_2})^{1/2}$, $1 \le l_1 \le l_2 \le L$.
\end{itemize}

Using this, the operating characteristics of a group sequential design with any choice of stopping boundaries can be determined using multivariate normal integration as described in \citet{jennison2000} and \citet{wason2015}. This allows the use of numerical optimisation routines to determine suitable sample sizes and stopping boundaries. The particular type of boundaries to utilise depends on the chosen hypothesis testing framework. Therefore, in the following sections we discuss several established methods for two-sided, and then one-sided, tests.

\section{Two-Sided Tests}

\subsection{Stopping Rules and Operating Characteristics}

In a two-sided test, we assess whether there is significant evidence of a difference in the mean responses of the two treatment arms. That is, we test
\[H_0 : \tau = 0, \qquad \qquad H_1 : \tau \neq 0.\]
Here, a group sequential trial design is characterised by stopping boundaries $a_1,\dots,a_L$ and $r_1,\dots,r_L$, with $0 \le a_l < r_l$ for $l=1,\dots,L-1$, and $a_L=r_L$, and the following stopping rules at analyses $l=1,\dots,L$
\begin{itemize}
	\item If $|Z_l| \ge r_l$ stop and reject $H_0$;
	\item If $|Z_l| < a_l$ stop and do not reject $H_0$;
	\item otherwise continue to stage $l+1$.
\end{itemize}
The choice $a_L=r_L$ ensures termination after analysis $L$, whilst also guaranteeing a conclusion is made about $H_0$.

Then, the probability of rejecting $H_0$ for any $\tau$, given $n$, is
\begin{align*}
P(\text{Reject }H_0 \mid \tau) &= \sum_{l=1}^L P(\text{Reject } H_0 \text{ at stage } l \mid \tau),\\
&= P(|Z_1| \ge r_1 \mid \tau)\\
& \qquad + \sum_{l=2}^L P(a_1 \le |Z_1| < r_1, \dots, a_{l-1} \le |Z_{l - 1}| < r_{l - 1},|Z_l| \ge r_l \mid \tau).
\end{align*}
Similarly, the probability of not rejecting $H_0$ for any $\tau$ is
\begin{align*}
P(\text{Accept } H_0 \mid \tau) &= \sum_{l=1}^L P(\text{Accept } H_0 \text{ at stage } l \mid \tau),\\
&= P(|Z_l| < a_l \mid \tau) \\
& \qquad + \sum_{l=2}^L P(a_1 \le |Z_1| < r_1, \dots, a_{l-1} \le |Z_{l - 1}| < r_{l - 1},|Z_l| < a_l \mid \tau).
\end{align*}
Using the above, the expected sample size for any $\tau$ can be calculated as
\[ E(N \mid \tau) = \sum_{l=1}^L \left\{P(\text{Reject } H_0 \text{ at stage } l \mid \tau) + P(\text{Accept } H_0 \text{ at stage } l \mid \tau)\right\}(ln + rln). \]

As discussed earlier, each of these probabilities can be computed using multivariate normal integration. Explicitly, defining
\begin{align*}
\Lambda_l &= \begin{pmatrix} cov(Z_1,Z_1) & \dots & cov(Z_1,Z_l) \\ \vdots & \ddots & \vdots \\ cov(Z_l,Z_1) & \dots & cov(Z_l,Z_l) \end{pmatrix},\\ 
\boldsymbol{I}_l &= (I_1,\dots,I_l)^T,
\end{align*}
then, for example
\begin{align*}
P(\text{Reject } H_0 \text{ at stage 2} \mid \tau) &= \int_{-r_1}^{-a_1}\int_{-\infty}^{-r_2} \phi\{\boldsymbol{x}, \tau \boldsymbol{I}_2^{1/2}, \Lambda_2\} \ \mathrm{d}x_2\mathrm{d}x_1 \\
& \qquad + \int_{-r_1}^{-a_1}\int_{r_2}^{\infty} \phi\{\boldsymbol{x}, \tau \boldsymbol{I}_2^{1/2}, \Lambda_2\} \ \mathrm{d}x_2\mathrm{d}x_1\\
& \qquad + \int_{a_1}^{r_1}\int_{-\infty}^{-r_2} \phi\{\boldsymbol{x}, \tau \boldsymbol{I}_2^{1/2}, \Lambda_2\} \ \mathrm{d}x_2\mathrm{d}x_1\\
& \qquad + \int_{a_1}^{r_1}\int_{r_2}^{\infty} \phi\{\boldsymbol{x}, \tau \boldsymbol{I}_2^{1/2}, \Lambda_2\} \ \mathrm{d}x_2\mathrm{d}x_1.
\end{align*}
Here, $\boldsymbol{x} = (x_1,x_2)^T$, the square root of the vector $\boldsymbol{I}_2$ is taken in an element wise manner, and \(\phi\{\boldsymbol{x},\boldsymbol{\mu},\Lambda\}\) is the probability density function of a multivariate normal distribution with mean \(\boldsymbol{\mu}\) and covariance matrix \(\Lambda\), evaluated at vector $\boldsymbol{x}$. In all of the commands presented here, these integrals are evaluated using the mata function \texttt{pmvnormal\_mata()} \citep{grayling2016}.

With the above specifications, all that remains is a method for determining stopping boundaries, and an associated required sample size, such that $P(\text{Reject } H_0 \mid 0)\le\alpha$ and $P(\text{Reject } H_0 \mid \pm \delta)\ge 1-\beta$, for clinically relevant difference $\delta>0$, and desired type-I and type-II error rates $\alpha$ and $\beta$. It is this problem that much of the group sequential clinical trial design literature has focused upon. In the following sections we discuss several options available via our commands.

\subsection{Early Stopping to Reject $H_0$}

Much of the early work on group sequential trial design focused on two-sided tests with early stopping only to reject $H_0$. That is, with $a_l=0$ for $l=1,\dots,L-1$. In particular, \citet{haybittle1971} and \citet{peto1976} suggested a simple set of boundaries with $r_l=3$ for $l=1,\dots,L-1$. The final critical boundary $r_L$ is then determined to ensure an overall type-I error rate of $\alpha$. Following the determination of $r_L$, a one-dimensional numerical search is utilised to ascertain the exact required group size $n$ for power of $1-\beta$ when $\tau=\pm\delta$, treating $n$ as a continuous quantity.

Haybittle and Peto's procedure is advantageous in that it is a simple one, whilst its wide stopping boundaries mean that early stopping is unlikely: a desirable property in some instances to help increase data accumulation, with termination only in the case of extreme disparities in treatment performance. However, trialists will often desire stopping boundaries that help to substantially reduce the expected sample size when $H_0$ is not true. For this, \citet{wang1987} suggested the following family of stopping boundaries, indexed by a parameter $\Omega$
\[r_l = C_{WT}(l/L)^{\Omega-1/2}.\]
Their procedure encompasses the popular \citet{pocock1977} and \citet{obrien1979} boundaries, by taking $\Omega=0.5$ or $\Omega=0$ respectively. In this approach, a numerical search is utilised for any chosen $\Omega$ to determine the value of $C_{WT}$ that implies the correct type-I error rate $\alpha$. Following this, as with Haybittle and Peto's design, a further search is then used to ascertain the required sample size for the power constraint. In general, it has been shown that as $\Omega$ increases, the maximum sample size increases, but the expected sample size for larger values of $|\tau|$ decreases. 

Later, we present commands \texttt{haybittlePeto} and \texttt{wangTsiatis} for determining the stopping boundaries and required sample size of these designs for any choice of $L$, $\sigma_0^2$, $\sigma_1^2$, $\delta$, $\alpha$, $\beta$ and $\Omega$. 

\subsection{Early Stopping to Reject and Not Reject $H_0$}

The above designs deal well with the issue of ethics in two-sided clinical trials; namely the desire to stop early when the difference between treatments is substantial. However, there are also often sound reasons to desire early stopping when it is clear there is no detectable treatment difference; usually based around reducing the cost of a trial. These are trial designs with not all $a_l=0$, $l=1,\dots,L-1$. \citet{pampallona1994} described a one-parameter family of such trial designs, again indexed by a shape parameter $\Omega$, that has been referred to as the power family of inner wedge designs. Explicitly
\begin{align*}
r_l &= C_{r}(l/L)^{\Omega-1/2},\\
a_l &= \delta I_l^{1/2} - C_{a}(l/L)^{\Omega-1/2}.
\end{align*} 
The final information level is then
\[ I_L = \frac{(C_{a} + C_{r})^2}{\delta^2},\]
to ensure $a_L=r_L$ as desired. A two-dimensional numerical search is utilised to determine the values of $C_{a}$ and $C_{r}$ that provide the desired type-I and type-II error rates given choices for $L$, $\alpha$, $\beta$ and $\Omega$. With these values identified, the final required information level $I_L$ is used to determine the exact required group size $n$ through Equation~(\ref{eq:I}). As in the procedure of \citet{wang1987} above, the inclusion of the parameter $\Omega$ allows a large range of designs to be determined, with varying performance in terms of their expected sample sizes. In Section~\ref{sec:ex1} we will see how these performances can be examined graphically.

Alternatively, \citet{whitehead1983} and \citet{whitehead1997} proposed an approach for the determination of a group sequential clinical trial design for a two-sided test with early stopping to not reject $H_0$, termed the double triangular test. Specifically, they demonstrated that a design with 
\begin{align*}
  r_l &= \left\{ \frac{2}{\tilde{\delta}}\log\left(\frac{1}{\alpha}\right) - 0.583\left(\frac{I_L}{L}\right)^{1/2} + \frac{\tilde{\delta}}{4} \frac{l}{L}I_L \right\}\frac{1}{I_l^{1/2}},\\
  a_l &= \text{max}\left\{\left[ -\frac{2}{\tilde{\delta}}\log\left(\frac{1}{\alpha}\right) + 0.583\left(\frac{I_L}{L}\right)^{1/2} + \frac{3\tilde{\delta}}{4} \frac{l}{L}I_L \right]\frac{1}{I_l^{1/2}}, 0 \right\},
\end{align*}
where
\[ \tilde{\delta} = \frac{2z_{1 - \alpha/2}\delta}{z_{1-\alpha/2} + z_{1 - \beta}},\]
and
\[ I_L = \left[\left\{\frac{4(0.583)^2}{L} + 8\log\left(\frac{1}{\alpha}\right)\right\}^{1/2} - \frac{2(0.583)}{L^{1/2}} \right]^2 \frac{1}{\tilde{\delta}} ,\]
would approximately attain a type-I error rate of $\alpha$ when $\tau=0$, and a type-II error rate of $\beta$ when $\tau=\pm\delta$.

Later, we discuss our commands \texttt{innerWedge} and \texttt{doubleTriangular} for determining these designs.

\section{One-Sided Tests} 

\subsection{Stopping Rules and Operating Characteristics}

In a one-sided test, we assess whether, without loss of generality, the mean response on treatment 1 is significantly larger than that on treatment 0. That is, we test
\[H_0 : \tau \le 0, \qquad \qquad H_1 : \tau > 0.\]
A group sequential trial design of this type is characterised by stopping boundaries $f_1,\dots,f_L$ and $e_1,\dots,e_L$, with $f_l < e_l$ for $l=1,\dots,L-1$ and $f_L=e_L$, and the following stopping rules at analyses $l=1,\dots,L$
\begin{itemize}
	\item If $Z_l \ge e_l$ stop and reject $H_0$,
	\item If $Z_l < f_l$ stop and do not reject $H_0$,
	\item otherwise continue to stage $l+1$,
\end{itemize}
Again, the choice $f_L=e_L$ is to ensure termination after analysis $L$, and to guarantee a conclusion is drawn about $H_0$.

Now, the probability of rejecting $H_0$ for any $\tau$, given $n$, becomes
\begin{align*}
P(\text{Reject } H_0 \mid \tau) &= \sum_{l=1}^L P(\text{Reject } H_0 \text{ at stage } l \mid \tau),\\
&= P(Z_l \ge e_l \mid \tau) \\
& \qquad + \sum_{l=2}^L P(f_1 \le Z_1 < e_1, \dots, f_{l-1} \le Z_{l - 1} < e_{l - 1},Z_l \ge e_l \mid \tau).
\end{align*}
Similarly, the probability of not rejecting $H_0$ for any $\tau$ is
\begin{align*}
P(\text{Accept } H_0 \mid \tau) &= \sum_{l=1}^L P(\text{Accept } H_0 \text{ at stage } l \mid \tau),\\
&=  P(Z_l < f_l \mid \tau) \\
& \qquad + \sum_{l=2}^L P(f_1 \le Z_1 < e_1, \dots, f_{l-1} \le Z_{l - 1} < e_{l - 1},Z_l < f_l \mid \tau).
\end{align*}
As before, the expected sample size for any $\tau$ is given by
\[ E(N \mid \tau) = \sum_{l=1}^L \left\{P(\text{Reject } H_0 \text{ at stage } l \mid \tau) + P(\text{Accept } H_0 \text{ at stage } l \mid \tau)\right\}(ln + rln). \]

Moreover, these probabilities can again be computed using multivariate normal integration. Using our notation from earlier, we have for example
\[P(\text{Reject } H_0 \text{ at stage } 2 \mid \tau) = \int_{f_1}^{e_1}\int_{e_2}^{\infty} \phi\{\boldsymbol{x}, \tau \boldsymbol{I}_2^{1/2}, \Sigma_2\} \ \mathrm{d}x_2\mathrm{d}x_1.\]

In some situations, a one-sided test will be more appropriate because departures from $H_0$ in one direction are implausible. Alternatively, it may be the case that we are interested in directly testing the superiority of one treatment over another. Consequently, much research has gone in to determining designs that will have desired operating characteristics (now, a type-I error rate of $\alpha$ when $\tau=0$, and a type-II error rate of $\beta$ when $\tau=\delta$) and favourable performance in terms of the expected sample size. Below, we discuss two popular methods, available for implementation via our commands.

\subsection{Power Family of One-Sided Designs}

In addition to their power family of inner wedge designs, \citet{pampallona1994} also detailed a one-parameter family of designs for one-sided tests, with boundaries given by
\begin{align*}
e_l &= C_{e}(l/L)^{\Delta-1/2},\\
f_l &= \delta I_l^{1/2} - C_{f}(l/L)^{\Omega-1/2}.
\end{align*} 
As before, taking a final information level of
\[ I_L = \frac{(C_{e} + C_{f})^2}{\delta^2},\]
ensures that $f_L=e_L$ as desired, and a two-dimensional grid search can be used to determine the appropriate values of $C_e$ and $C_f$. Our command \texttt{powerFamily} is available to perform these computations.

\subsection{Triangular Test}

\citet{whitehead1983} and \citet{whitehead1997} also proposed a triangular test for one-sided group sequential clinical trial designs. Specifically, they proposed
\begin{align*}
e_l &= \left\{ \frac{2}{\tilde{\delta}}\log\left(\frac{1}{2\alpha}\right) - 0.583\left(\frac{I_L}{L}\right)^{1/2} + \frac{\tilde{\delta}}{4} \frac{l}{L}I_L \right\}\frac{1}{I_l^{1/2}},\\
f_l &= \left\{ -\frac{2}{\tilde{\delta}}\log\left(\frac{1}{2\alpha}\right) + 0.583\left(\frac{I_L}{L}\right)^{1/2} + \frac{3\tilde{\delta}}{4} \frac{l}{L}I_L \right\}\frac{1}{I_l^{1/2}},
\end{align*}
with
\[ \tilde{\delta} = \frac{2z_{1 - \alpha/2}\delta}{z_{1-\alpha/2} + z_{1 - \beta}},\]
and
\[ I_L = \left[\left\{\frac{4(0.583)^2}{L} + 8\log\left(\frac{1}{2\alpha}\right)\right\}^{1/2} - \frac{2(0.583)}{L^{1/2}} \right]^2 \frac{1}{\tilde{\delta}},\]
demonstrating this design would approximately attain the desired operating characteristics.

This design has proven popular with trialists because of the speed with which it can be calculated, and also because of its strong performance in terms of its expected sample sizes \citep{wason2012}. Our command \texttt{triangular} determines this design.

\section{Syntax}

In this section, we detail the syntax of our six discussed commands, which are all declared as \texttt{rclass}

\begin{stsyntax}
	doubleTriangular, \optional{l(integer 3) \underbar{d}elta(real 0.2) \underbar{a}lpha(real 0.05) \underbar{b}eta(real 0.2) \underbar{s}igma(numlist) \underbar{r}atio(real 1) \underbar{perf}ormance *}
\end{stsyntax}

\begin{stsyntax}
	haybittlePeto, \optional{l(integer 3) \underbar{d}elta(real 0.2) \underbar{a}lpha(real 0.05) \underbar{b}eta(real 0.2) \underbar{s}igma(numlist) \underbar{r}atio(real 1) \underbar{perf}ormance *}
\end{stsyntax}

\begin{stsyntax}
	innerWedge, \optional{l(integer 3) \underbar{d}elta(real 0.2) \underbar{a}lpha(real 0.05) \underbar{b}eta(real 0.2) \underbar{s}igma(numlist) \underbar{r}atio(real 1) \underbar{o}mega(real 0.5) \underbar{perf}ormance *}
\end{stsyntax}

\begin{stsyntax}
	powerFamily, \optional{l(integer 3) \underbar{d}elta(real 0.2) \underbar{a}lpha(real 0.05) \underbar{b}eta(real 0.2) \underbar{s}igma(numlist) \underbar{r}atio(real 1) \underbar{o}mega(real 0.5) \underbar{perf}ormance *}
\end{stsyntax}

\begin{stsyntax}
	triangular, \optional{l(integer 3) \underbar{d}elta(real 0.2) \underbar{a}lpha(real 0.05) \underbar{b}eta(real 0.2) \underbar{s}igma(numlist) \underbar{r}atio(real 1) \underbar{perf}ormance *}
\end{stsyntax}

\begin{stsyntax}
	wangTsiatis, \optional{l(integer 3) \underbar{d}elta(real 0.2) \underbar{a}lpha(real 0.05) \underbar{b}eta(real 0.2) \underbar{s}igma(numlist) \underbar{r}atio(real 1) \underbar{o}mega(real 0.5) \underbar{perf}ormance *}
\end{stsyntax}

Here, the prescribed options denote the following

\hangpara{\texttt{\underbar{a}lpha} is the desired overall type-I error rate, $\alpha$. That is, it is the two-sided or one-sided type-I error rate according to the chosen command.}

\hangpara{\texttt{\underbar{b}eta} is the desired type-II error rate, $\beta$.}

\hangpara{\texttt{\underbar{d}elta} is the clinically relevant difference at which we power, $\delta$.}

\hangpara{\texttt{l} is the maximum number of allowed stages in the design, $L$.}

\hangpara{\texttt{\underbar{o}mega} is the shape parameter of the boundaries of the power family and Wang-Tsiatis designs.}

\hangpara{\texttt{\underbar{perf}ormance} specifies that the performance of the identified design, i.e. its expected sample size and power curves, should be determined and plotted.}

\hangpara{\texttt{\underbar{r}atio} is the desired ratio $r$ of the sample sizes between arms 0 and 1.}

\hangpara{\texttt{\underbar{s}igma} is the standard deviation of the responses in arms 0 and 1; $\sigma_0$ and $\sigma_1$. This can either be of length two, containing the assumed values of these two parameters, or of length one, implying $\sigma_0=\sigma_1$.}

Attainable via \texttt{return list} for all six commands, are the determined exact required group size $n$ (\texttt{r(n)}), and the stopping boundaries $a$, $r$, $e$ and $f$ as appropriate (e.g., \texttt{r(a)}). In addition, the vector of information levels $\boldsymbol{I}$ (\texttt{r(I)}), the covariance matrix $\Lambda$ (\texttt{r(Lambda)}), and a vector summarising the performance of the design (\texttt{r(performance)})
\[(P(\text{Reject } H_0 \mid 0), E(N\mid 0),P(\text{Reject } H_0 \mid \delta),E(N \mid \delta),\text{max}_\tau E(N\mid\tau), \text{max} N)^T,\]
are available.

Note that in all of these commands, required one dimensional numerical searches are performed using a purpose built implementation of Brent's algorithm \citep{Brent1973}. In contrast, all two dimensional numerical searches are carried out with the Nelder-Mead option in \texttt{optimize()}.

\section{Example 1: Two-Sided Tests}\label{sec:ex1}

As our first example, we consider the case $L=2$, $\alpha=0.05$, $\beta=0.2$, $\delta=0.2$, $\sigma_0=\sigma_1=2$, and $r=1$, in two-sided testing.

We begin by demonstrating how \texttt{doubleTriangular} can be used to determine the boundaries and sample size required by the Double Triangular test of \citet{whitehead1983}. Explicitly, the following code is used to determine the design

\begin{stlog}
	. doubleTriangular, l(2) alpha(0.05) beta(0.2) delta(0.2) sigma(2) r(1)
{\smallskip}
2-stage Group Sequential Trial Design
\HLI{37}
The hypotheses to be tested are as follows:
{\smallskip}
  H0: tau = 0  H1: tau != 0,
{\smallskip}
with the following error constraints:
{\smallskip}
  P(Reject H0 | tau = 0) = .05,
  P(Reject H0 | tau = delta = .2) = 1 - .2.
{\smallskip}
   Double-triangular boundaries selected....................
...now determining design...................................
...design determined. Returning the results.................
...Exact required group sizes for each arm determined to be:
{\smallskip}
  875.5 and 875.5.
{\smallskip}
...Rejection boundaries r determined to be:
{\smallskip}
  (2.2,2.07).
{\smallskip}
...Acceptance boundaries a determined to be:
{\smallskip}
  (.73,2.07).
{\smallskip}
...Operating characteristics of the design are:
{\smallskip}
  P(Reject H0 | tau = 0)  = .0531,
  P(Reject H0 | tau = .2) = .8003,
  E(N | tau = 0)          = 2514.6,
  E(N | tau = .2)         = 2550.5,
  max_tau E(N | tau)      = 2716.4,
  max N                   = 3501.9.
\end{stlog}

As can be seen, by default the commands return an informative summary of the chosen testing framework, their progress, and the characteristics of the final design. Specifically,  the first few lines describe the hypotheses that will be tested based on the chosen command. The input values of \texttt{alpha} and \texttt{beta} are then used in printing a summary of the desired operating characteristics. Several lines then follow which describe the progress of the command in completing its required computations. Next, the exact required number of patients in each arm, in each stage, are printed. The rejection and acceptance boundaries then follow, along with a summary of the operating characteristics of the identified design. In this case we see the design has a type-I error-rate of 0.053, and power of 0.800. This is a well-known limitation of the double triangular design: the type-I and type-II error requirements are only approximately achieved. The final four printed results summarise various important sample size characteristics of the design: the expected sample size when $\tau=0$, that when $\tau=\delta$, the maximum expected sample size over all possible values of $\tau$, and the maximum possible required sample size. We can see that in this case, whilst the maximum possible value of $N$ is 3501.9, we would expected to not require more than 2716.4 patients.

Being able to easily determine this design is useful, however in most situations it is unlikely that a trialist will have a single design in mind. Consequently, it is important to be able to determine the performance of several designs and compare them graphically. Here, we demonstrate this for the power family of inner wedge designs. Using the following code, we find the designs for $\Omega=-0.5$, $-0.25$, $0$ and $0.25$, saving their performance. Then, we combine the saved graphs to produce Figure 1

\begin{stlog}
	. qui innerWedge, l(2) alpha(0.05) beta(0.2) delta(0.2) sigma(2) omega(-0.5) r(1)
> perf saving(firstDesign) nodraw title({\lbr}\&Omega{\rbr} = -0.5) scale(0.75) scheme(sj)
{\smallskip}
. qui innerWedge, l(2) alpha(0.05) beta(0.2) delta(0.2) sigma(2) omega(-0.25) r(1)
> perf saving(secondDesign) no draw title({\lbr}\&Omega{\rbr} = -0.25) scale(0.75) scheme(sj)
{\smallskip}
. qui innerWedge, l(2) alpha(0.05) beta(0.2) delta(0.2) sigma(2) omega(0) r(1) perf
> saving(thirdDesign) nodraw title({\lbr}\&Omega{\rbr} = 0) scale(0.75) scheme(sj)
{\smallskip}
. qui innerWedge, l(2) alpha(0.05) beta(0.2) delta(0.2) sigma(2) omega(0.25) r(1)
> perf saving(fourthDesign) nodraw title({\lbr}\&Omega{\rbr} = 0.25) scale(0.75) scheme(sj)
{\smallskip}
. graph combine firstDesign.gph secondDesign.gph thirdDesign.gph fourthDesign.gph,
> ycommon scheme(sj)

\end{stlog}

\begin{figure}[hbtp]
	\centering
	\label{fig1}
	\includegraphics[width=0.9\textwidth]{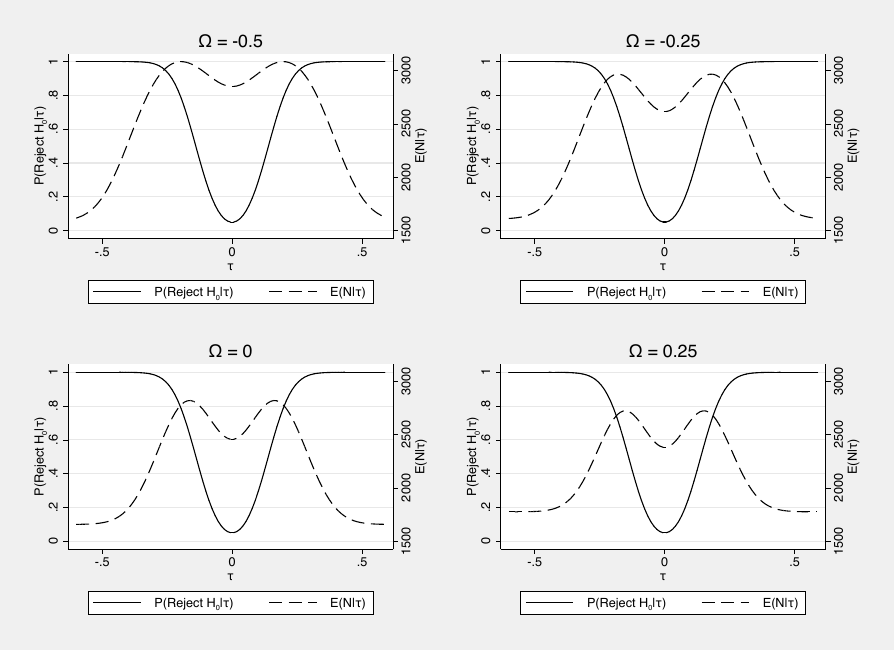}
	\caption{Comparison of the performance of several two-sided power family of inner wedge designs.}
\end{figure}

We observe that increasing the value of $\Omega$ appears to reduce the expected sample required when $|\tau|$ is small. However, this comes at a cost to that required when $|\tau|$ is large.

\section{Example 2: One-Sided Tests}

As our next example, we consider one-sided testing. We take $L=3$, $\alpha=0.1$, $\beta=0.1$, $\delta=0.25$, $\sigma_0=1$, $\sigma_1=2$ and $r=2$. Similarly to the above, we demonstrate how \texttt{powerFamily} can be used to determine several designs ($\Omega=-0.25$, $\Omega=0$, and $\Omega=0.25$), and in addition compute the boundaries and sample size of the triangular test. Saving the performance of each, we then compare their performance graphically, creating Figure 2 with the following code

\begin{stlog}
	. qui powerFamily, l(3) alpha(0.1) beta(0.1) delta(0.25) sigma(1, 2) omega(-0.25)
> r(2) perf saving(firstDesign) nodraw title(Power family with {\lbr}\&Omega{\rbr} = -0.25)
> scale(0.75) scheme(sj)
{\smallskip}
. qui powerFamily, l(3) alpha(0.1) beta(0.1) delta(0.25) sigma(1, 2) omega(0) r(2)
> perf saving(secondDesign) nodraw title(Power family with {\lbr}\&Omega{\rbr} = 0)
> scale(0.75) scheme(sj)
{\smallskip}
. qui powerFamily, l(3) alpha(0.1) beta(0.1) delta(0.25) sigma(1, 2) omega(0.25)
> r(2) perf saving(thirdDesign) nodraw title(Power family with {\lbr}\&Omega{\rbr} = 0.25)
> scale(0.75) scheme(sj)
{\smallskip}
. qui triangular, l(3) alpha(0.1) beta(0.1) delta(0.25) sigma(1, 2) r(2) perf 
> saving(fourthDesign) nodraw title(Triangular test) scale(0.75) scheme(sj)
{\smallskip}
. graph combine firstDesign.gph secondDesign.gph thirdDesign.gph fourthDesign.gph,
> ycommon scheme(sj)

\end{stlog}

\begin{figure}[hbtp]
	\centering
	\label{fig2}
	\includegraphics[width=0.95\textwidth]{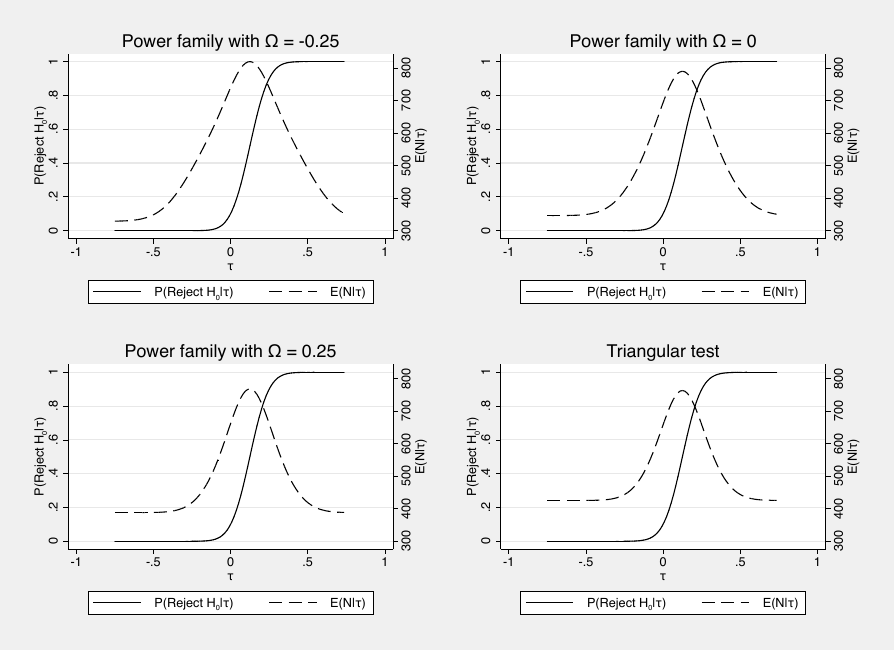}
	\caption{Comparison of the performance of several one-sided power family designs and the triangular test.}
\end{figure}

As has been reported previously, the triangular test does indeed fare well in comparison to the two identified power family designs. Explicitly, it has the lowest maximum expected sample size of the four designs. However, this does come at the cost of an increased maximum possible sample size, as evidence by its performance for large $|\tau|$.

\section{Conclusion}

It is important that any clinical trial control both its type-I and type-II error rates accurately. For this task, Stata introduced in Version 13 the command \texttt{power}, which can be used for an extremely wide array of trial scenarios. However, as we have discussed, group sequential clinical trial designs are extremely popular with researchers, and to date few available commands are available in Stata for determining such designs. Notable exceptions include \texttt{nstage} \citep{Barthel2009,Bratton2015} and \texttt{nstagebin} \citep{Bratton2014} for multi-arm multi-stage trial designs with time-to-event and binary endpoints respectively. In addition, the command \texttt{simsam} can determine the required sample size of certain group sequential clinical trial designs given stopping boundaries \citep{Hooper2013}. There are no established commands however for determining the boundaries and group size required by the wide array of group sequential trial designs for normally distributed outcomes discussed here.

Several extensions to our commands are now possible. We have assumed that the variance of the responses on both treatment arms is known prior to trial commencement. Whilst this is a common assumption in the group sequential design literature, often this will be a strong one to make. However, \citet{whitehead2009} proposed a simple quantile substitution method for dealing with this problem, which has been shown to generally control the type-I error rate to the correct level \citep{wason2012a}. This would no doubt be a useful addition to our commands. Moreover, we have assumed that the interim analyses are equally spaced in-terms of the number of patient responses accrued in each arm. \citet{lan1983} proposed an error spending approach to the design of group sequential trials that allows this assumption to be relaxed. Consequently, a command to employ such methodology could prove useful to those seeking more complex designs.

Additionally, our focus has been on two-arm trials. Today, multi-arm multi-stage trials are becoming increasingly popular. Therefore, extending these designs to allow for multiple experimental arms would be advantageous. Finally, there have now been several proposals for the determination of optimal or near-optimal group sequential designs (see, for example, \citet{wason2012}, \citet{wason2012a}, and \citet{Wason2015a}). To allow trialists to maximise the efficiency gains made by utilising a group sequential design, the establishment of commands for determining such designs would be highly advantageous.

Regardless of these possible expansions, our commands can be used to determine stopping boundaries, exact required group sizes, and also to compare the performance of a selection of designs. Consequently, they should prove useful to those seeking to exploit the efficiencies of a group sequential design whilst working in Stata.

\section{Acknowledgements}

Michael J. Grayling is supported by the Wellcome Trust (Grant Number 099770/Z/12/Z). James M. S. Wason is supported by the National Institute for Health Research Cambridge Biomedical Research Centre (Grant Number MC\_UP\_1302/6). Adrian P. Mander is supported by the Medical Research Council (Grant Number MC\_UP\_1302/2).

\bibliographystyle{sj}
\bibliography{sj}

\begin{aboutauthors}
Michael J. Grayling is an Investigator Statistician at the Medical Research Council Biostatistics Unit in Cambridge, UK.

James M. S. Wason is a Programme Leader Track at the Medical Research Council Biostatistics Unit in Cambridge, UK.

Adrian P. Mander is the Director of the Hub for Trials Methodology Research at the Medical Research Council Biostatistics Unit in Cambridge, UK.
\end{aboutauthors}

\clearpage
\end{document}